\makeatletter
\def\input@path{{accessibility/}}
\makeatother
\DocumentMetadata{
  lang=en-US,
  testphase={phase-III,table},
  pdfstandard=ua-1
}
\documentclass[manuscript]{acmart}

\setcopyright{none}
\renewcommand\footnotetextcopyrightpermission[1]{}

\usepackage{booktabs}
\usepackage{tabularx}
\usepackage{longtable}
\usepackage{array}
\usepackage{calc}
\usepackage{graphicx}
\graphicspath{{figures/}{./}}
\providecommand{\tightlist}{%
  \setlength{\itemsep}{0pt}\setlength{\parskip}{0pt}}

\begin{document}

\title{Beyond Prompt-to-App: Accountable Translation in Teacher-Facing Agentic Authoring}

\author{Nizam Kadir}
\email{nizam_kadir@mymail.sutd.edu.sg}
\affiliation{%
  \department{Science, Mathematics and Technology}
  \institution{Singapore University of Technology and Design}
  \city{Singapore}
  \country{Singapore}}

\author{Wei Ting Liow}
\email{weiting_liow@sutd.edu.sg}
\affiliation{%
  \department{Science, Mathematics and Technology}
  \institution{Singapore University of Technology and Design}
  \city{Singapore}
  \country{Singapore}}

\author{Sumbul Khan}
\email{sumbul_khan@sutd.edu.sg}
\affiliation{%
  \department{Science, Mathematics and Technology}
  \institution{Singapore University of Technology and Design}
  \city{Singapore}
  \country{Singapore}}

\author{Lay Kee Ang}
\email{ricky_ang@sutd.edu.sg}
\affiliation{%
  \department{Science, Mathematics and Technology}
  \institution{Singapore University of Technology and Design}
  \city{Singapore}
  \country{Singapore}}

\renewcommand{\shortauthors}{Kadir et al.}

\begin{abstract}
Natural-language app builders let domain experts create software, but their pipelines transform professional intent across compilation, generation, checking, and approval. We report a bounded trace study of a teacher-facing agentic authoring system. Evidence comprises six eligible build attempts across three accounts; a separate corpus of 37 workshop units from 23 display names contextualizes commitments without person-level linkage. Compiled specifications added governance requirements, while downstream representations sometimes normalized case-specific learning relations. Two drafts met a stored package/security threshold despite analyzer reservations and unresolved correspondence to their briefs; four attempts in one account produced no usable payload, and repair messages did not translate internal terms into domain-legible revisions. We develop accountable translation as an analytic framework for making consequential changes attributable, inspectable, scoped in validation, and contestable. It extends HCI accounts of traceability and end-user debugging by locating professional authority and repair rights across heterogeneous technical and organizational handoffs.
\end{abstract}

\keywords{agentic authoring, end-user development, educational AI, teacher agency, natural-language programming, human--AI interaction, validation, repairability, traceability}

\ccsdesc[500]{Human-centered computing~Empirical studies in HCI}
\ccsdesc[500]{Human-centered computing~HCI theory, concepts and models}
\ccsdesc[300]{Applied computing~Education}
\ccsdesc[300]{Software and its engineering~End-user software engineering}

\maketitle

\hypersetup{
  pdfauthor={Nizam Kadir; Wei Ting Liow; Sumbul Khan; Lay Kee Ang},
  pdftitle={Beyond Prompt-to-App: Accountable Translation in Teacher-Facing Agentic Authoring},
  pdfsubject={Human-centered computing; empirical studies in HCI; HCI theory and models; education; end-user software engineering}
}

\providecommand{\tagpdfsetup}[1]{}
\section{Introduction}\label{introduction}

Natural-language programming appears to remove a longstanding barrier to end-user development: instead of learning syntax, a domain expert can describe an application and ask an AI system to produce it. This shift is especially attractive in education. Educators routinely encounter situated problems that generic software does not address: helping learners articulate misconceptions, preserving evidence of reasoning, differentiating feedback, or coordinating a classroom routine. A teacher-facing app builder could make those problems directly authorable rather than waiting for a product team to generalize them.

Removing programming syntax, however, does not remove specification work. A short brief still has to become a data model, interaction sequence, interface, evidence policy, runtime package, and reviewable artifact. The system must decide what the user meant, what platform capabilities can realize that meaning, and what constraints should take priority. These decisions are consequential when generated software may later mediate student work. A fluent preview can preserve the topic while changing the learning activity; a technically valid package can omit the teacher's intended control; and a failed build can name an internal contract term without explaining how the teacher should revise the design. Natural language therefore relocates rather than eliminates the work of software specification.

We study this relocation through a teacher-facing agentic authoring system that turns a natural-language educational brief into a planned, generated, checked, and reviewable app package. We call the system \textbf{Studio} throughout. Studio uses several model- and rule-based stages rather than a single prompt-response call. A compiler expands the brief with platform contracts; model stages plan and generate the application; automated checks inspect package and security requirements; and a registry gate separates a draft from software that is available to others. This pipeline provides a concrete setting in which to ask not merely whether output was generated, but what happened to explicit professional commitments across heterogeneous handoffs.

Our empirical setting was a three-hour professional-learning and structured-ideation workshop with educators. The primary analysis traces six eligible Studio attempts across three accounts from brief through compiled specification, plan, static artifact, analyzer output, package check, and registry state. A separate contextual public corpus contains 37 participant-authored ideation units, including problems of practice, group framings, critical questions, and structured app proposals. We use that corpus only to sensitize the analysis to purposes, controls, evidence, and safeguards expressed during the workshop. We do not claim that a public idea belongs to a Studio case, that a display name is a verified person, or that the two sources form a participation funnel.

We use \textbf{translation} for the empirical process through which one representation becomes another. \textbf{Accountable translation} is the normative property at stake: consequential changes to purpose, decision rights, evidence and data boundaries, validity scope, and repair rights should be attributable to a stage, inspectable by the domain author, supported by evidence of stated scope, and contestable through domain-legible repair. A \textbf{professional commitment} here is an explicit authored actor--action, control, evidence, data, safeguard, or success relation in the retained text; it is not a claim about private intent or pedagogical endorsement. In the contextual public corpus, many units articulated bounded scaffolding, feedback, diagnosis, and process evidence, although controls, evidence flows, and safeguards were unevenly visible in posted specifications. In the Studio traces, compiled specifications introduced governance-related requirements, while later app-family assignments, plans, and artifacts normalized distinctive learning relations. Local completion and package signals coexisted with analyzer reservations, static correspondence gaps, unusable payloads, and no observed registry entries.

These observations lead to three research questions:

\textbf{RQ1.} How did explicit professional commitments change as the eligible Studio briefs moved through compilation, planning, generation, checking, repair, and registry state?

\textbf{RQ2.} Where did stage-level status claims, analyst-assessed brief correspondence, repair guidance, and registry availability diverge?

\textbf{RQ3.} What design propositions for accountable translation follow from these observed handoffs?

This paper contributes three things. First, we provide a cross-stage empirical account of how governance requirements accumulated, case-specific learning relations were sometimes normalized, and local status signals remained fragmented across six eligible attempts. Second, we develop \textbf{accountable translation} as an analytic framework whose unit is a professional commitment---and the authority, evidence scope, and repair path attached to it---as it moves through stochastic and organizational handoffs. Third, we derive design propositions for semantic provenance, scoped validation, and domain-legible repair. The study offers a representation- and handoff-oriented account of agentic authoring, not estimates of prevalence or evidence of classroom effectiveness.

\section{Related Work}\label{related-work}

\subsection{End-user development after natural-language generation}\label{end-user-development-after-natural-language-generation}

End-user programming and end-user development aim to let people create or modify computational artifacts without becoming professional programmers \citep{lieberman2006eud, myers2006eup}. Natural-programming research similarly asks how computational representations can match the ways people express intent \citep{myers2004natural}. Their central challenge has never been syntax alone. Domain experts still perform requirements, testing, debugging, risk management, and integration work \citep{ko2011euse}. Natural-language generation changes the interface to this work, but it does not dissolve the underlying need for specification and debugging.

Generative interfaces can make software production resemble conversation, but natural-language specification remains difficult. Users discover an implicit model-facing ``syntax,'' while concrete test cases can improve task communication where clarification alone may not \citep{jiang2022nlp,pickering2025communicate}. Code-assistant studies distinguish accelerating a known implementation from exploring an uncertain one \citep{vaithilingam2022expectation,barke2023grounded}. Recent systems expose this mediation: Pail tracks confirmed requirements and LLM-introduced design decisions; Jelly and semantic-guided UI generation use inspectable intermediate representations; and SimStep makes task-level abstractions editable for educational-simulation authoring \citep{zamfirescu2025pail,cao2025jelly,park2026semantic,kaputa2026simstep}. They address important local transformations, while leaving open responsibility, evidence scope, and authorization across a full app pipeline.

Agentic app authoring intensifies the problem because output is executable and persistent: screens, storage, permissions, analytics, error states, and downstream users. Prompt-based tools can accelerate high-fidelity prototypes across skill levels, while deployment, security, and maintainability still depend on technical expertise \citep{kobiella2026throwaway}. A successful response is therefore not a usable or available artifact. We shift end-user development from access to code toward accountability across translation.

\subsection{Delegation, oversight, and appropriate reliance}\label{delegation-oversight-and-appropriate-reliance}

Human--AI interaction research has long argued that automation should support appropriate reliance rather than maximal delegation. Trust is calibrated when people can understand capabilities, observe state, intervene, and recognize uncertainty \citep{lee2004trust, shneiderman2020hcai}. Systems that infer intent require intelligible interpretations, action feedback, and retained control \citep{bellotti2001intelligibility}. Delegability varies with motivation, difficulty, risk, and trust \citep{lubars2019delegability}; interaction guidance and recent oversight work also require visible state, efficient correction, legible responsibility, and meaningful human contribution \citep{amershi2019guidelines,faas2026oversight}. These principles become operational questions in a pipeline: Which stage inferred a requirement? What did a template add? What did a validator inspect? What remains editable?

Agent chains offer one response by representing an AI task as inspectable stages \citep{wu2022aichains}. Recent multi-agent debugging work pairs overview navigation with failure localization, prior-message editing, and execution reset \citep{epperson2025debugging}. Yet decomposition does not make handoffs accountable. A stage can complete with malformed output; a later checker can pass a package while an earlier reviewer withholds approval; and a final status can collapse divergent signals. Confidence cues can also increase acceptance of correct and incorrect recommendations \citep{bansal2021whole}. The HCI problem is therefore not only stage visibility but \textbf{handoff observability}: whether the interface shows what meaning, authority, and validity changed between stages.

This concern connects to requirements traceability in software engineering. Traceability links stakeholder needs to implementation and verification evidence \citep{gotel1994traceability,clelandhuang2014traceability}; recent retrieval-augmented work extends link recovery across heterogeneous artifacts while retaining practical limits \citep{fuchss2025lissa}. Recovered links are technical evidence, not endorsement of a semantic change. In AI-generated software, the relevant question is whether an intended learning relation---such as questioning without answering---survives app-family selection, planning, generation, and checking. Automated repair also depends on an oracle that defines acceptable behavior \citep{legoues2019repair}; package tests can serve syntax and contract checks, but not by themselves alignment with a domain requirement. We therefore distinguish package compliance from analyst-assessed brief correspondence.

\subsection{Teacher agency and classroom orchestration}\label{teacher-agency-and-classroom-orchestration}

Teachers' work is a strong case for studying accountable translation because educational decisions are situated, relational, and institutionally constrained. Classroom orchestration research describes how teachers coordinate people, materials, timing, evidence, and interventions under real-time constraints \citep{dillenbourg2013orchestration}. Studies of teacher routines show that classroom technologies must fit existing professional practices rather than treat teaching as an abstract sequence of tasks \citep{an2017routines}. Work on teacher--AI complementarity similarly argues that AI systems should make their analyses actionable within teachers' orchestration needs and preserve meaningful opportunities for professional judgment \citep{holstein2019codesigning, holstein2019complementarity}.

Teacher agency is not simply the presence of an approval button. It involves the capacity to shape goals, interpret evidence, respond to context, and act within institutional conditions \citep{imants2020agency}. Multi-year teacher-centered work on AI co-orchestration likewise shows that shared control implicates trust, responsibility, efficiency, and accuracy rather than a binary human-versus-AI choice \citep{lawrence2024sharedcontrol}. Educational AI therefore raises questions about who may formulate a pedagogical purpose, decide what reaches learners, interpret uncertain evidence, or authorize data use. Human oversight matters because teachers remain responsible for consequences that an opaque generator may not represent \citep{molenaar2022hybrid, nguyen2023ethical}.

Participatory and collaborative design can make this professional knowledge visible. Teacher talk and co-design artifacts can reveal classroom routines, curricular constraints, and tacit criteria that would otherwise remain outside a system specification \citep{mckenney2016codesign, sperling2024behind}. However, eliciting a good idea is only the first part of authoring. Once a professional problem is handed to an agentic development pipeline, its intermediate representations may preserve or transform the relation encoded in that idea. Our deliberately non-linked sources address the two sides separately: public posts document an intent vocabulary, while a small eligible Studio subset documents pipeline transformations.

\subsection{From prompt quality to accountable translation}\label{from-prompt-quality-to-accountable-translation}

Prior work provides principles for end-user development, inspectable AI, and teacher agency, but the combination exposes a gap. Prompt-oriented accounts tend to evaluate whether a user can elicit a desired output. Education-oriented accounts tend to focus on how a teacher uses or oversees an already-existing AI capability. Neither perspective fully captures a pipeline in which an educator's language becomes software that must be compiled, checked, repaired, approved, and made available.

We use \textbf{accountable translation} as a normative property, not a synonym for the pipeline. Translation is accountable when consequential changes to (1) purpose and interaction, (2) decision rights and professional controls, (3) evidence and data boundaries, (4) validity scope, and (5) repair rights are attributable, inspectable, supported by scoped evidence, and contestable in domain language. This procedural view resonates with contestability work that locates meaningful challenge in sensemaking, responsibility attribution, and accessible sites of intervention rather than a final explanation alone \citep{yurrita2025contestability}. The prompt is the first representation in a chain that remains answerable to its motivating professional commitments.

Accountable translation is not another name for traceability, chain visibility, debugging, or teacher agency. Traceability connects requirements to artifacts and tests; accountable translation asks who can alter a professional commitment, what evidence warrants the change, and how it can be challenged. Visibility reveals stages but does not assign responsibility or define what a status supports. Debugging provides means to correct an artifact; accountable translation also follows changes to decision rights, data practices, and organizational approval. Teacher-agency research explains why professional judgment matters; our framework traces where that judgment is reassigned or restored during authoring. Its novelty lies in treating a professional commitment, together with its authority, evidentiary scope, and repair path, as the unit that must remain answerable across technical and organizational handoffs.

This distinction is consequential even in relation to the closest authoring systems. Pail can preserve a requirement while recording an introduced design decision, and SimStep can expose staged teacher-facing representations for inspection, refinement, and inverse correction \citep{zamfirescu2025pail,kaputa2026simstep}. Yet a trace link may coexist with a changed actor--action relation; a refinable representation may still leave unclear what evidence a status supports; and a repair control may stop before organizational review or availability. Accountable translation extends these interaction patterns by binding representation change to professional authority, explicitly scoped validity, and a contestation path across both technical and organizational gates.

\section{System Context: A Governed Agentic Authoring Pipeline}\label{system-context-a-governed-agentic-authoring-pipeline}

Studio is a no-code authoring environment in which an educator describes a proposed learning application in natural language. The system does not return a single free-standing answer. It creates a task that moves through planning, generation, analysis, deterministic checking, and review states. An artifact may then be previewed and, after additional review, entered into a plugin registry. Registry availability is separate from build completion.

A prior system preprint describes the intended architecture of the underlying platform: a prepared brief can move through generation, preview, safety checking, administrator review, directory publication, and gated telemetry \citep{kadir2026studio}. That paper documents an intended architecture and demonstrated lifecycle, not the workshop traces analyzed here. The present paper is a separate empirical study of eligible, naturally occurring attempts, including blocked and unresolved outcomes. We use the preprint only for system provenance, not as evidence that the current cases executed, passed review, or became available.

Before model work begins, the system expands the brief into a compiled prompt. The compiled prompt includes platform runtime constraints, allowed capabilities, accessibility and error-state requirements, storage and telemetry rules, review expectations, and an app-family template. Model stages then produce a plan, build specification, application payload, and analysis. Deterministic checks inspect package structure, security, and runtime-contract conditions. The pipeline stores intermediate inputs and outputs, artifact status, a numeric package/security score, and selected research-event records.

This architecture intentionally introduces governance-related requirements into a short request. For example, the compiled specification can require administrator review before distribution, disallow unsupported functions, constrain storage, and request empty, loading, and error states. Their presence in a specification does not establish that a generated artifact enforces them. At the same time, the pipeline chooses how to normalize an account holder's idea: an app-family assignment, named purpose, and generic success criteria shape later plans. The mechanism that makes generation governable can therefore also transform the domain problem.

At the study snapshot, the deployed prototype combined a deterministic compiler and package validator with service-routed model stages for planning, generation, and analysis. Provider selection, inference settings, and a complete version history were not preserved in the de-identified case extracts in a form suitable for model-level attribution. We therefore analyze observable differences between stored representations rather than attribute outcomes to a named model or isolate a software cause; provider, prompt, template, routing, hidden state, and validator changes remain alternative explanations and replication variables.

\begin{figure*}[t]
  \centering
  \includegraphics[alt={Four de-identified actual Studio interface screenshots show a prepared brief and deterministic preflight, staged agent handoffs, a sandboxed sample preview, and scoped validation with human review. Product-brand labels are replaced with generic platform wording; the prepared content is not participant data.},width=\textwidth]{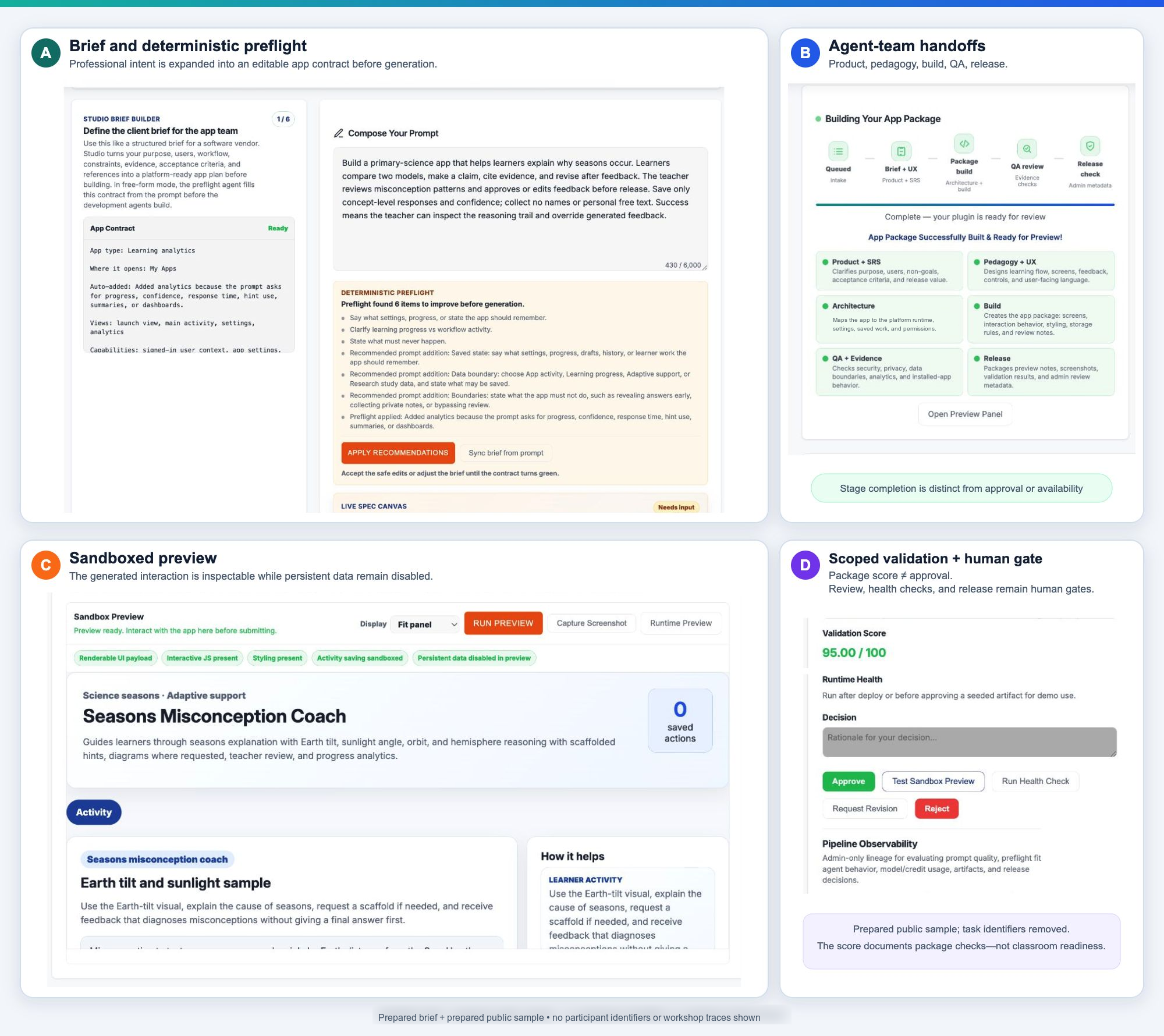}
  \caption{De-identified actual Studio authoring interfaces captured after the workshop from a later prepared build and populated with a non-participant science brief and sample application. Product-brand labels were deterministically replaced with generic platform wording; the underlying UI composition is otherwise unchanged. (A) Preflight expands a brief into an editable contract. (B) The interface exposes product, pedagogy, build, quality, and release handoffs. (C) A generated interaction is inspected in a sandbox. (D) Package validation, runtime health, and human approval are distinct. The displayed content and score document system surfaces, not a participant-visible study state, workshop trace, study outcome, or classroom-readiness claim.}
  \Description{Four de-identified actual Studio interface screenshots are arranged in panels. Panel A shows a prepared brief and deterministic preflight. Panel B shows staged handoffs from intake through release checking. Panel C shows a sandboxed prepared sample preview. Panel D separates a package score from human review, runtime health, and approval. Product-brand labels are deterministically replaced with generic platform wording, and no participant content is shown.}
  \label{fig:workflow}
\end{figure*}

\begin{figure*}[t]
  \centering
  \includegraphics[alt={A de-identified actual generated-application interface for a prepared science sample has six numbered callouts identifying professional purpose, problem structure, a scientific visual model, layered learner support, professional control, and inspectable state. Product-brand labels are replaced with generic platform wording; no participant content is shown.},width=\textwidth,height=0.68\textheight,keepaspectratio]{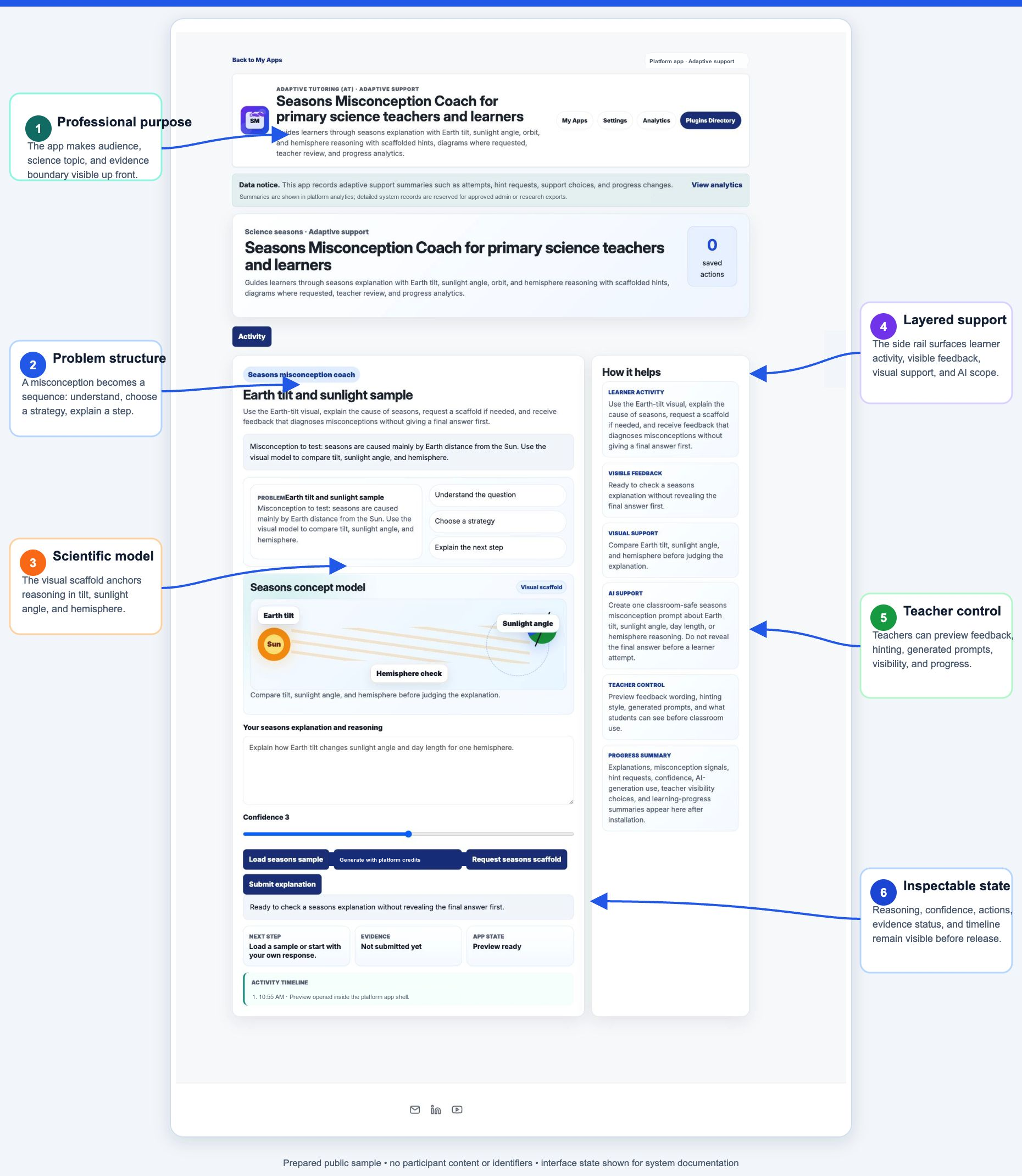}
  \caption{De-identified actual generated-application interface captured after the workshop from a later prepared build for a non-participant science sample. Product-brand labels were deterministically replaced with generic platform wording; the underlying UI composition is otherwise unchanged. The annotations identify professional purpose, problem structure, a visual model, layered learner support, professional control, and inspectable state. The screenshot documents later output and governance surfaces in Studio; it is not a participant artifact, participant-visible study state, workshop trace, or evidence of runtime behavior, account-holder acceptance, classroom use, or learning effects.}
  \Description{A de-identified actual generated-application interface for a prepared sample is surrounded by six numbered callouts. They identify professional purpose, problem structure, a scientific visual model, layered learner support, professional control, and inspectable state. Product-brand labels are deterministically replaced with generic platform wording, and no participant content or workshop trace is shown.}
  \label{fig:sample-app}
\end{figure*}

Figures~\ref{fig:workflow} and~\ref{fig:sample-app} document actual Studio surfaces captured after the workshop from a later prepared build. They are included to make the authoring and governance interfaces legible; they do not establish which interface state an account holder saw during the study. The empirical trace evidence begins with the retained records described in Method.

\section{Method}\label{method}

\subsection{Bounded trace study and setting}\label{bounded-trace-study-and-setting}

We conducted a bounded, multi-source qualitative trace study during a three-hour professional-learning and structured-ideation workshop in mid-2026. The session invited educators to articulate problems of practice, formulate possible educational applications, and encounter a deployed agentic authoring prototype. It was not a controlled usability study, classroom deployment, or evaluation of learning outcomes.

The recorded attendance roster contained 29 attendee records from 10 schools. Seventeen attendees reported at least ten years of teaching experience. Because teaching level was a multi-select field, selections exceed the attendance count: 18 selected Lower Secondary, 20 Upper Secondary, and three Junior College. Roster records were not linked to public-chat display names or Studio accounts and do not define the analytic sample.

The session moved from problem framing to structured application proposals. An eight-part scaffold elicited an app name, intended users, problem, function or AI action, professional control, data or evidence, risks or safeguards, and an indicator of success. Studio use occurred during the same session, but public contributions and Studio accounts were not linked. Access and form problems may have affected who reached Studio and how long they spent authoring, so the eligible Studio attempts are not treated as representative of workshop attendance.

\subsection{Data sources and inclusion boundaries}\label{data-sources-and-inclusion-boundaries}

Table~\ref{tab:evidence-ledger} distinguishes the parallel records and their denominators. The first analytic source is a participant-authored public workshop corpus. We reconstructed the earlier corpus segment from a preserved rich capture and the later segment from the official public-chat export; a fixed handoff within their overlap prevented duplicate inclusion. Submission-facing materials label these coarse segments \texttt{early\_discussion} and \texttt{later\_public\_chat}, while the exact boundary remains in the restricted co-author audit archive. Within the retained records, exact duplicates were removed, while two consecutive posts that completed one structured proposal were merged as one conceptual unit. We excluded private or direct messages, participant codes, administration, help exchanges, names and identifying details, reactions as substantive data, facilitator-authored conceptual content, brief acknowledgments, and timing messages. The resulting corpus comprised 37 substantive participant-authored units from 23 display names: 10 problem-of-practice units, six group app framings, three critical questions, and 18 later structured ideas. Display names are not treated as verified people or attendance.

The primary source is a de-identified Studio trace subset. At metadata level, seven same-window, check-in-linked tasks were considered. Inclusion then used stored research-participation and interaction-log eligibility flags; build outcome was not an inclusion rule. Six build attempts from three accounts were eligible and are labelled T1--T3. One additional task lacked the relevant interaction-log flag and was excluded without opening its content. The de-identified extracts do not preserve enough provenance to establish exactly when or how every flag was assigned, which limits independent auditing of selection.

For eligible attempts we examined the account-authored portion of the prompt, compiled prompt, generated name and description, plan, build specification, prompt history, task-step inputs and outputs, selected user-facing artifact strings, statically declared or referenced runtime APIs, validator and static-analysis reports, research-event state, and registry state at the study snapshot. We did not reproduce raw prompts, full code, identifiers, or secrets, and we did not execute the generated apps.

\begin{table}[tbp]
\small
\centering
\caption{Evidence ledger for parallel, non-linkable records. Counts use each source's own unit and do not form an attrition sequence.}
\label{tab:evidence-ledger}
\tagpdfsetup{table/header-rows={1}}
\setlength{\tabcolsep}{3pt}
\begin{tabularx}{\linewidth}{@{}>{\raggedright\arraybackslash}p{0.22\linewidth}>{\raggedright\arraybackslash}p{0.17\linewidth}>{\raggedright\arraybackslash}p{0.20\linewidth}>{\raggedright\arraybackslash}X@{}}
\hline
Record source & Unit & Retained record & Analytic role and boundary \\
\hline
Attendance roster & Attendee record & 29 records; 10 schools & Aggregate setting context only; no linkage to analytic traces \\
Operational check-in & Platform record & 26 records & Workshop operations only; not verified attendance or sample size \\
Public workshop corpus & Substantive unit / display name & 37 units; 23 display names & Contextual vocabulary of expressed purposes, controls, evidence, and safeguards \\
Structured-idea subset & Public unit & 18 of 37 units & Explicit-field coverage after the scaffold; not individual change \\
Studio metadata screen & Task & 7 tasks & Eligibility screening only \\
Eligible Studio traces & Attempt / account & 6 attempts; 3 accounts & Primary trace analysis; T1=1, T2=1, T3=4 dependent attempts \\
Excluded Studio task & Task & 1 task & Content unopened because the relevant flag was absent \\
Registry snapshot & Eligible artifact entry & 0 observed & Availability at one snapshot; publication attempt not established \\
\hline
\end{tabularx}
\Description{A parallel evidence ledger distinguishes attendance records, operational check-ins, public units, structured ideas, Studio tasks and attempts, an excluded task, and registry state. The rows are not a participant funnel.}
\end{table}

Additional planned workshop forms yielded no analyzable downstream responses and are not used. Reaction marks are retained only as contextual provenance; because facilitator reactions made up nearly half and unique reactors cannot be inferred, we do not use reactions as preferences, votes, or outcome evidence.

\subsection{Analysis}\label{analysis}

The public-corpus unit was one retained substantive participant-authored contribution after applying the source boundary, deduplication, merge, and exclusion rules described above. We assigned four mutually exclusive unit types and nine non-exclusive descriptive theme families. A unit could receive multiple theme codes when its text explicitly addressed more than one concern. The code families index the preserved record; they are not latent constructs, participant attributes, or prevalence estimates.

Coding proceeded in two aggregate-aware passes. A research-team analyst with prior knowledge of the system and workshop segmented the corpus and produced the initial unit types, theme totals, and structured-field totals. Codex then recoded the units against the operational definitions while the first-pass aggregate totals were visible. Unit-level first-pass assignments had not been preserved, so the passes were neither independent nor suitable for intercoder-reliability statistics. Seven of nine theme totals and all six structured-field totals initially matched. A later record-level source audit corrected four rows (U04, U12, U22, and U25); the audited matrix, codebook, and aggregate checker report the final decisions. The research team defined the code boundaries and remains responsible for them. We treat the values as model-assisted descriptive indexing, not evidence of coding reliability or bias reduction.

The assisted pass used OpenAI Codex in the desktop research environment in September 2026; an auditable model-build identifier and complete prompt transcript were not preserved. The model received de-identified analytic units and the codebook definitions, not names, schools, participant codes, Studio identifiers, or full raw prompts. Instructions required one mutually exclusive unit type, any explicitly supported non-exclusive themes and structured fields, a bounded paraphrase and rationale, preservation of source separation, and no inference of latent intent. Discrepancies were resolved by clarifying the inclusive code boundaries documented in the supplement and regenerating the matrix; there was no blinded adjudicator. The local analytic record preserves the resulting rows and arithmetic checks, but not service-side retention settings. These omissions limit procedural replication and reinforce our treatment of the pass as non-independent analytic assistance.

For Studio, the comparison unit was an explicit account-authored commitment nested within an attempt and account. We reconstructed an ordered evidence chain across the account brief, compiled specification, app-family assignment, plan or build specification, inspected static artifact, analyzer message, package/security check, and registry snapshot. Platform-introduced requirements were recorded separately from account-authored commitments.

At each stage, we coded a commitment as \emph{preserved} when the same actor--action relation remained recognizable; \emph{transformed} when a related element remained but its actor, interaction, decision right, or evidence relation changed; \emph{not found in inspected static material} when the bounded search and manual inspection located no representation; or \emph{unresolved} when the retained evidence could not support a judgment. ``Not found'' does not imply runtime absence. A de-identified requirement-by-stage matrix reports a bounded stage summary and rationale for each decision.

An insider research-team analyst conducted the Studio correspondence assessment using the retained traces and documented system knowledge. T1 and T2 were analyzed as individual attempts and T3-1--T3-4 as one dependent within-account sequence for negative-case analysis. Stored pipeline state, package/security state, analyst-assessed static correspondence, and registry presence were recorded as separate evidence scopes. These assessments do not establish account-holder endorsement, runtime behavior, pedagogical effectiveness, or classroom validity.

The anonymized supplement provides a 37-row public-corpus matrix, a 33-row Studio requirement trace, the six-attempt joint display, code definitions, decision rationales, and a deterministic aggregate-consistency check. The matrices record the reported analytic decisions and bounded stage summaries; the script checks transcription and arithmetic, not source-to-result reproduction. Neither the matrices nor the script recover the missing first-pass unit assignments or constitute independent reliability evidence.

Integration was abductive and conceptual. The contextual public corpus supplied a vocabulary of professional commitments, while the Studio traces exposed transformations in three separate eligible accounts. We organized recurring commitments and trace differences into five accountability objects: purpose and interaction, decision rights, evidence and data boundaries, validity scope, and repair rights. We then related these objects to traceability, oversight, debugging, and contestability to articulate four normative properties: attributable change, inspectability, explicitly scoped validation, and contestable repair. These properties are an analytic synthesis rather than a measured scale. The cross-stage joint display juxtaposes the two sources but is not participant-level triangulation. We do not attribute any public idea to T1--T3, infer that the scaffold caused individual learning, or generalize frequencies to educators as a population.

\subsection{Ethics, anonymity, and positionality}\label{ethics-anonymity-and-positionality}

The workshop and the de-identified, model-assisted analyses reported here were covered by institutional ethics review and an approved protocol amendment. Participation in research was voluntary and separable from participation in the professional-learning activity. We included only public contributions and Studio traces within the approved scope and relevant consent settings. To limit re-identification, the manuscript omits the protocol identifier, system URLs, participant names and codes, schools, and raw data. The Codex-assisted pass used de-identified analytic units rather than identifying records or full prompts. Short public excerpts were minimized and context was removed to reduce, but not eliminate, re-identification risk.

The research team included members with system-development and workshop-facilitation roles. These insider positions provided access to intermediate traces that are ordinarily hidden, but they also create confirmation risk and may shape what counts as an important requirement or mismatch. Retaining negative cases, separating the two sources, stating the model-assisted coding procedure, and distinguishing local status scopes make those interpretive decisions more visible; they do not constitute independent corroboration or remove bias. The analysis is intended to examine consequential transformations in a system to which the team had insider access, not to demonstrate platform success.

\section{Results}\label{results}

\subsection{Professional commitments in a contextual public corpus}\label{professional-commitments-in-a-contextual-public-corpus}

The public corpus is not a precursor linked to the Studio cases; it contextualizes the concerns visible in workshop posts. The most frequently coded units concerned learner agency or process scaffolding (21/37), feedback, diagnosis, or differentiated support (18/37), risks/safeguards (16/37), evidence/data (15/37), and workload or instructional orchestration (13/37). Only five centered content or activity generation. Because codes overlapped and units combined individual, group-reported, and phase-specific contributions, these figures describe posted-unit coverage rather than participant prevalence or consensus. Table~\ref{tab:themes} reports the overlapping descriptive themes and one illustrative unit for each inclusion decision.

Many posts bounded what an app should do in relation to learner work. Examples included questioning assumptions without supplying finished answers, preserving reasoning and revision histories, and returning class patterns for professional interpretation. One short phrase requested ``guiding suggestions rather than simply giving students the answers'' (E1); another proposed flags ``without automatically judging misconduct'' (E2). E1 and E2 are neutral excerpt labels from the public corpus and are not linked to Studio. Counterexamples included scheduling, a deliberately playful coercive-device idea, and proposals with little governance. The corpus therefore establishes a vocabulary of possible professional commitments, not a uniformly mature set of specifications.

\begin{table}[tbp]
\small
\centering
\caption{Overlapping themes in 37 participant-authored public units. Counts refer to units, not people; examples illustrate code inclusion rather than typicality. Quoted phrases are short excerpts and other examples are paraphrases.}
\label{tab:themes}
\tagpdfsetup{table/header-rows={1}}
\setlength{\tabcolsep}{3pt}
\begin{tabularx}{\linewidth}{@{}>{\raggedright\arraybackslash}p{0.34\linewidth}r>{\raggedright\arraybackslash}X@{}}
\hline
Theme family & $n$ & Illustrative unit \\
\hline
Learner agency / process scaffolding & 21 & ``guiding suggestions rather than simply giving students the answers'' \\
Feedback, diagnosis, or differentiated support & 18 & Prior-knowledge questions used to surface misconceptions and class patterns (paraphrase) \\
Data / evidence / analytics & 15 & Comparison of initial and revised essays (paraphrase) \\
Risks / safeguards & 16 & ``without automatically judging misconduct'' \\
Workload or instructional orchestration & 13 & A class-level summary of common errors and strengths (paraphrase) \\
AI literacy / authenticity & 11 & Drafts, sources, revisions, and declared AI use (paraphrase) \\
Teacher control / oversight & 10 & ``Teacher reviews before sharing'' \\
Interface / access / adoption & 7 & A question about whether generated apps resemble chatbots and what interfaces are possible (paraphrase) \\
Content / activity generation & 5 & Generation of differentiated questions or activities (paraphrase) \\
\hline
\end{tabularx}
\Description{Nine overlapping theme families are listed with counts and one short illustrative unit each.}
\end{table}

\subsection{Posted-specification coverage after the scaffold}\label{posted-specification-coverage-after-the-scaffold}

Table~\ref{tab:structured-fields} reports what was explicit in the 18 retained structured ideas. All named users and a function, 16 named a learning or operational problem, 12 named data or evidence, 10 named professional control, and nine named a risk or safeguard. These counts describe the preserved posts. An omitted field may have been discussed but not posted, and the record cannot determine whether omission arose from time, channel compression, group reporting, the scaffold, or conceptual uncertainty.

When present, controls were concrete: ``Teacher reviews before sharing'' and ``Teachers make the final judgement.'' Evidence was framed as class patterns, revision histories, or visible reasoning rather than only system usage. Safeguards included privacy, bias and accuracy checks, non-judgmental flags, and restrictions on direct student exposure. These fields supplied the sensitizing concepts used to compare Studio representations; they are not attributed to any Studio account.

\begin{table}[tbp]
\small
\centering
\caption{Fields explicitly visible in 18 posted structured ideas. Omission from a retained post does not establish absence from the underlying group deliberation.}
\label{tab:structured-fields}
\tagpdfsetup{table/header-rows={1}}
\begin{tabularx}{0.78\linewidth}{@{}>{\raggedright\arraybackslash}Xr@{}}
\hline
Specification element & Structured ideas (N=18) \\
\hline
Intended users explicit & 18 \\
Function or purpose explicit & 18 \\
Learning or operational problem explicit & 16 \\
Teacher control / oversight explicit & 10 \\
Data collection or evidence returned explicit & 12 \\
Risk or safeguard explicit & 9 \\
\hline
\end{tabularx}
\Description{Six specification elements are listed for 18 structured ideas: users and function appear in all 18, a problem in 16, evidence in 12, professional control in 10, and a safeguard in nine.}
\end{table}

The sequence marks only a change in the \emph{form of elicited output}. Only four of the eight contributors to the early problem phase also posted a structured idea, so it cannot support an individual learning claim. The contextual corpus instead sensitizes the trace analysis to whether professional authority, evidence requirements, and safeguards remain explicit after a brief enters the pipeline.

\subsection{Compilation added governance terms and reshaped case-specific interactions}\label{compilation-added-governance-terms-and-reshaped-case-specific-interactions}

The Studio traces begin with three recognizable domain problems. T1 requested a five-part workflow in which learners would document progress and the AI would question assumptions without supplying answers. T2 requested an authorship and critical-thinking companion addressing plagiarism or AI contribution, fair judgment, and individualized dialogue. T3 requested an upper-secondary science revision environment combining educator-provided outcomes and resources, generated practice, spaced repetition, misconception-sensitive feedback, mastery tracking, and a revision schedule. T3 later reformulated the brief to make account roles, resource upload, question editing, and analytics more explicit.

Studio expanded each brief into a compiled prompt of approximately 11,000--12,000 characters. The account-authored portion represented 2.0\%--7.5\% of the resulting prompt by character count. This ratio does not measure causal influence, importance, or attention: a short requirement can determine a large implementation. It only documents representational asymmetry. The brief entered a larger platform-authored specification naming runtime capabilities, storage and telemetry rules, accessibility expectations, security constraints, error states, evidence profiles, and review gates.

This \textbf{governance accretion} introduced governance-related requirements into every compiled specification: bounded storage, human-readable analytics, loading and error states, accessibility, review before distribution, and prohibitions on unsupported or unsafe functions. The traces establish the textual presence of those requirements, not their runtime enforcement. The pipeline also assigned an app family and generic success criteria, such as opening an installed app, completing an interaction, observing a state change, and saving approved activity.

At later stages, the same pipeline normalized case-specific intent. T1 and T3 were assigned to ``adaptive orchestration''; T2 was assigned to ``retrieval practice.'' T1's recognizable five-part workflow became a generic hint-and-progress sequence in the plan and inspected artifact. T2's originality and dialogic-feedback problem became a one-prompt-at-a-time recall workflow. T3's resource-grounded, subject-specific revision environment became a generic learner-attempt/hint/progress sequence. The evidence locates these changes across app-family assignment, plan, and artifact; it does not establish that the compiler alone caused them.

Stored ``clarification answered'' events did not provide evidence of account-holder dialogue. They appeared almost immediately after submission and contained system-produced answers about app family, capabilities, reviewer, and telemetry. We therefore treat them as system-generated specification records. The observation is about stored provenance: an event label can imply human clarification even when ambiguity was resolved internally.

Table~\ref{tab:attempts} preserves attempt-level dependence rather than collapsing four T3 attempts into one outcome. Timing is relative within T3; no comparison across accounts is intended.

\begin{table}[tbp]
\scriptsize
\centering
\caption{Cross-case display of input or attempt condition, focal representation change, downstream status, and available repair for six eligible Studio attempts. T3-1--T3-4 are dependent attempts from one account. Package/security scores reproduce checker-specific stored values; static correspondence does not establish runtime behavior or author endorsement.}
\label{tab:attempts}
\tagpdfsetup{table/header-rows={1}}
\setlength{\tabcolsep}{2pt}
\begin{tabularx}{\linewidth}{@{}>{\raggedright\arraybackslash}p{0.07\linewidth}*{4}{>{\raggedright\arraybackslash}X}@{}}
\hline
Attempt & Input or attempt condition & Focal representation change & Downstream representation and status & Available repair and evidence boundary \\
\hline
T1 & Five-part workflow; question assumptions; do not supply answers & Adaptive-orchestration family assignment & Hints and non-answering remained; requested workflow not found; valid JSON analyzer record with QA manual-review note; regex-enhanced package/security score 95 against minimum 60; no registry entry & Score calibration, runtime, brief alignment, human review, and availability unverified \\
T2 & Originality/AI contribution; fair judgment; critical dialogue & Retrieval-practice family assignment & Related title remained; flashcard recall with AI generation disabled; valid JSON analyzer record with QA manual-review note; score 95 against minimum 60; no registry entry & No retained domain-level revision; score calibration, runtime, and account-holder satisfaction unverified \\
T3-1 & Detailed science-revision environment & Adaptive-orchestration plan & No usable payload; score 0; `passed=false'; `package\_structure'; ``reflection journal'' message & Label did not state a domain-level change \\
T3-2 & Exact same detailed brief, 164 seconds later & Same compiled brief, plan, and required-capability set & Same coder record and no-payload failure class; score 0; `passed=false' & Dependent retry; no causal attribution or proposed domain change \\
T3-3 & Structured reformulation after 23 minutes 45 seconds & Extracted capabilities changed & File upload and AI generation explicit; no payload; score 0; `passed=false'; `package\_structure'; ``artifact review'' message & Label did not state a domain-level change \\
T3-4 & Exact revised text, 119 seconds later & Extracted capabilities changed again & File upload retained; AI generation absent; same no-payload failure class; score 0; `passed=false' & Identical text yielded a different representation; no causal attribution or proposed domain change \\
\hline
\end{tabularx}
\Description{Six rows compare each input or attempt condition, focal representation change, downstream representation and status, and available repair with evidence boundaries. T1 and T2 have a regex-enhanced package/security score of 95 against a minimum threshold of 60. The four dependent T3 attempts have score zero, a false passed flag, package-structure category, and no usable payload.}
\end{table}

\begin{figure*}[t]
  \centering
  \includegraphics[alt={Seven boxes trace T2 from an originality and dialogic-feedback brief through compilation, retrieval-practice planning, a flashcard-recall static artifact, a valid JSON analyzer record with an embedded manual-review QA note, a regex-enhanced package/security score of 95 against a minimum threshold of 60, and registry absence.},width=\textwidth]{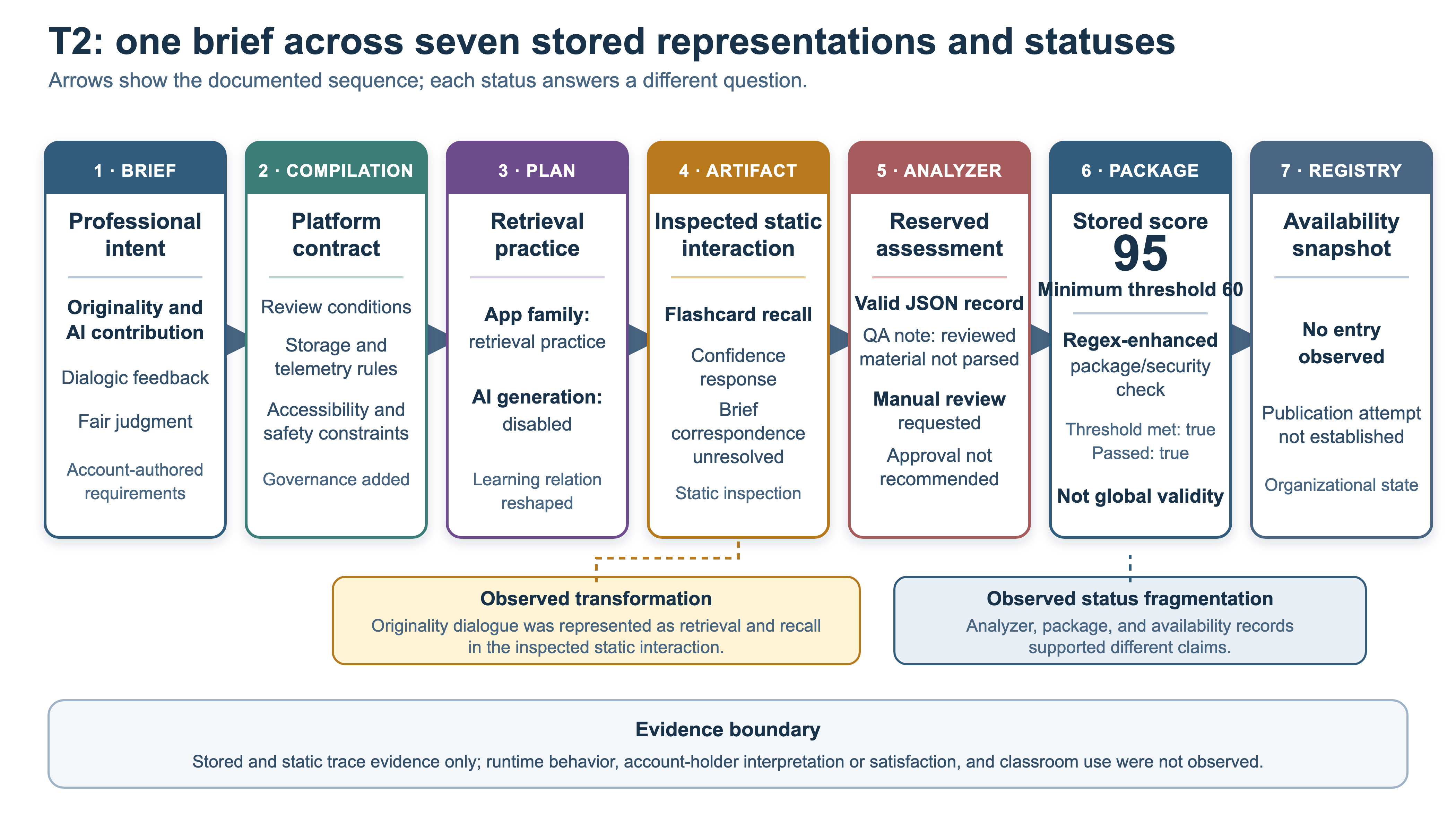}
  \caption{Cross-stage trace for T2. An originality- and dialogic-feedback brief was expanded with platform governance, classified as retrieval practice, and represented in the inspected static artifact as flashcard recall. The stored analyzer record was valid JSON; its QA note requested manual review. A later regex-enhanced package/security check stored score 95 against minimum threshold 60; no registry entry was observed. Each status is reported within its evidentiary scope.}
  \Description{Seven left-to-right boxes trace T2 from the account-authored brief through compilation, planning, the inspected static artifact, analyzer output, package checking, and registry state. The brief requests originality and AI-contribution support, dialogic feedback, and fair judgment. Compilation adds review, storage, telemetry, accessibility, and safety conditions. The plan assigns retrieval practice and disables AI generation. The inspected artifact centers flashcard recall. The analyzer record is valid JSON; an embedded QA note says reviewed material could not be parsed, flags manual review, and does not recommend approval. The regex-enhanced package and security check stores score 95 against a minimum threshold of 60. No registry entry is observed. Callouts identify a change in the represented interaction and different scopes among analyzer, package, and availability records. A footer limits claims to stored and static trace evidence.}
  \label{fig:t2-trace}
\end{figure*}

Figure~\ref{fig:t2-trace} makes the T2 sequence visible without collapsing the scopes of its stored records. It is a trace visualization assembled from the bounded evidence described in Method, not a participant-facing screen or runtime observation.

\subsection{Stored status layers answered different questions}\label{stored-status-layers-answered-different-questions}

The six attempts produced several locally valid but non-equivalent status records. All 16 recorded task steps reached a terminal ledger state. The planner-research record stored an expected-JSON parse note for T2 and all four T3 attempts (five of six); T1's planner record parsed. Separately, the T1 and T2 analyzer records were valid JSON, but their embedded QA notes said the reviewed material could not be parsed, requested manual review, and did not recommend approval. T1 and T2 produced payloads whose regex-enhanced package/security records stored score 95 against minimum threshold 60, with `meets\_threshold=true' and `passed=true'. The four T3 attempts produced no usable payload and stored score 0, `passed=false', and category `package\_structure'. No eligible artifact was present in the registry at the snapshot. These counts have different units and evidentiary scopes; they do not form a success rate.

T1 and T2 make the distinction concrete. Both regex-enhanced package/security checks stored score 95 against minimum threshold 60, yet static comparison located only partial correspondence to their explicit briefs. T1 retained hints and a non-answering boundary while the requested five-part workflow was not found. T2 retained a related title while the inspected activity became recall and confidence reporting. The evidence supports divergence among stored representations and status scopes, not claims about score calibration, runtime behavior, what account holders noticed, or whether they regarded an artifact as acceptable.

The four scopes are: \textbf{pipeline-step state}, which records whether a step terminated and persisted something; \textbf{package/security state}, which records the implementation-specific output or score of automated package/security checks; \textbf{analyst-assessed static correspondence}, which compares explicit brief commitments with inspected material; and \textbf{registry state}, which reports availability at one organizational snapshot. A numeric score from one layer cannot certify the others.

\subsection{Stored repair messages used platform categories rather than proposed domain changes}\label{stored-repair-messages-used-platform-categories-rather-than-proposed-domain-changes}

T3 provides a dependent within-account sequence of revision and repeat submission. The first two attempts used the same detailed science-revision brief 164 seconds apart. Their submitted text, compiled brief, implementation plan, required-capability set, coder record, and no-payload failure class were byte-identical. Both stored score 0, `passed=false', and category `package\_structure'. After 23 minutes 45 seconds, the account holder reformulated the idea into a shorter, structured brief that explicitly named users, the learning problem, application actions, professional controls, and class or student analytics. A fourth attempt repeated that revised text 119 seconds later. Both revised attempts also produced no usable payload and the same package-state values. The traces do not establish the account holder's motive for either repetition or a causal mechanism.

The reformulation changed the stored representation. File upload became explicit, and controls and analytics were easier to identify. One revised attempt included file upload and AI generation; the exact-text retry retained file upload but omitted AI generation even though question generation remained central. Identical submitted text therefore did not yield an identical stored capability plan. Because provider, inference, routing, version, and hidden-state information was not preserved, the trace cannot isolate why.

Across T3-1--T3-4, the coder withheld malformed or unsafe dynamic payloads and issued refunds. Separate fidelity messages invoked ``reflection journal'' for the first pair and ``artifact review'' for the revised pair. The trace does not show that these terms caused the package failure or that the account holder read them. It shows that the stored messages did not connect the internal label to a proposed learner action, evidence relation, or editable change in the account brief. The finding concerns message content, not user comprehension or repair experience.

\subsection{Cross-stage synthesis: accountability was fragmented across representations}\label{cross-stage-synthesis-accountability-was-fragmented-across-representations}

Table~\ref{tab:synthesis} juxtaposes the contextual public corpus and Studio transformations at the level of accountability objects only. It does not link a public contributor to T1--T3. The pattern is parsimonious: a brief entered a larger platform contract; governance requirements accrued; an app-family and plan reshaped the interaction; local checks reported status within heterogeneous scopes; and repair messages did not translate internal terms into proposed changes to the account holder's specification.

\begin{table}[tbp]
\scriptsize
\centering
\caption{Accountability objects across the separate public-corpus and Studio sources, bounded empirical consequences, and corresponding design questions. Rows do not link participants across sources.}
\label{tab:synthesis}
\tagpdfsetup{table/header-rows={1}}
\setlength{\tabcolsep}{2.5pt}
\begin{tabularx}{\linewidth}{@{}*{5}{>{\raggedright\arraybackslash}X}@{}}
\hline
Accountability object & Contextual public corpus & Studio trace & Supported inference & Design question \\
\hline
Purpose and interaction & 21/37 process-scaffolding units; 18/37 feedback, diagnosis, or support & T1--T3 were assigned generic families; T1/T2 static interactions differed from explicit briefs & Topic or title can persist while the actor--action relation changes & How can authors compare brief, contract, plan, and artifact? \\
Decision rights & 10/18 structured ideas named professional control & Compiled prompts introduced review language; inspected evidence did not consistently expose the requested control & A textual review rule does not establish an implemented or visible control & Where can an author inspect, edit, or reject a decision right? \\
Evidence and data boundaries & 12/18 named evidence; 9/18 named safeguards & Compilation added storage and telemetry requirements; static API references varied & Requested evidence, permitted collection, static implementation, and approval are different claims & How should the interface separate these states? \\
Validity scope & --- & 16/16 steps terminal; 2/6 score 95 against minimum 60; 5/6 planner parse notes; 4/6 no payload; 0 registry entries & A local signal cannot certify end-to-end readiness & What evidence and residual uncertainty should accompany each status? \\
Repair rights & --- & Four dependent T3 attempts included reformulation and exact retries; stored messages used ``reflection journal'' or ``artifact review'' without stating a domain-level change & The available message did not translate the reported contract term into an editable specification & How can repair connect the original statement, missing interaction, and proposed change? \\
\hline
\end{tabularx}
\Description{Five accountability objects connect available public-corpus counts and bounded Studio observations to supported inferences and design questions. Dashes indicate that the public corpus did not contain the relevant trace evidence. Rows are conceptual and are not person-level links.}
\end{table}

The registry contained no entry for any eligible artifact at the study snapshot. We do not know whether publication was attempted, whether sufficient review time had elapsed, or whether the absence was an expected consequence of the manual gate. The defensible outcome is therefore narrower than ``zero successful apps'': two payloads stored regex-enhanced package/security scores of 95 against minimum threshold 60, four attempts produced no usable payload, and no eligible artifact was observed in the registry. This is neither a success-rate estimate nor evidence of checker calibration, runtime behavior, or classroom readiness.

The traces also do not show what ordinary users understood from the interface. They show a \textbf{cross-stage status fragmentation} in stored evidence: local completion and package signals coexisted with analyzer reservations, static correspondence gaps, failures, and registry absence. Whether this fragmentation was participant-visible, confusing, or consequential for reliance remains a question for direct interface and user evaluation.

\section{Discussion}\label{discussion}

The empirical contribution is not that generated applications sometimes fail. Rather, the traces show how responsibility became distributed across representations that carried different evidentiary scopes. The compiler added governance conditions; application-family assignment and planning sometimes normalized the proposed learning relation; package, analyzer, correspondence, and registry records answered different questions. No single stage accounts for the whole outcome, yet the domain author would ultimately be expected to stand behind the resulting application. Accountable translation names the need to keep that chain answerable to the professional commitments with which it began.

\begin{figure*}[t]
  \centering
  \includegraphics[alt={A separate contextual public corpus sits beside an observed Studio chain from account brief through compilation and agent work to four evidence scopes: pipeline-step state, package and security state, analyst-assessed static correspondence, and registry state; the requested manual artifact review was not evidenced in the retained trace. Three callouts identify interaction reshaping, terminal status with a planner parse note and repair need, and package score as distinct from readiness or registry presence. A proposed lower overlay contains four accountability tests: attributable change, inspectable handoffs, scoped validity, and contestable repair.},width=\textwidth]{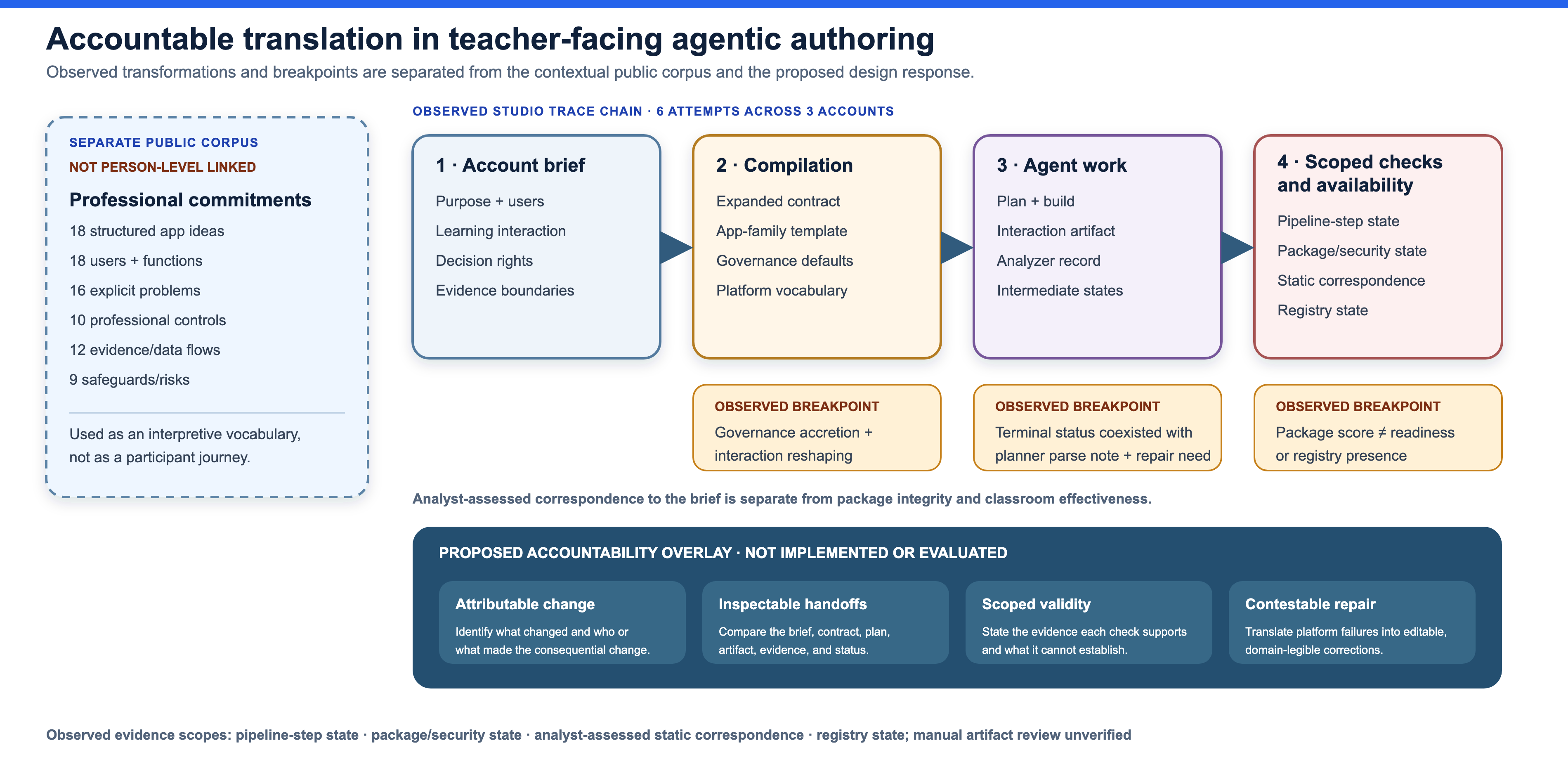}
  \caption{Accountable-translation model used in the analysis. A separate public corpus contextualizes professional commitments; the observed Studio chain follows eligible representations and statuses; the lower layer specifies four proposed accountability tests: attributable change, inspectable handoffs, scoped validity, and contestable repair.}
  \Description{A separate and unlinked contextual public corpus sits beside an observed Studio chain. The chain runs from an account brief through platform compilation and agent work to four evidence scopes: pipeline-step state, package and security state, analyst-assessed static correspondence, and registry state; the requested manual artifact review was not evidenced in the retained trace. Callouts identify governance accretion with interaction reshaping, terminal status coexisting with a planner parse note and repair need, and a package score that is distinct from readiness or registry presence. A lower banner specifies attributable change, inspectable handoffs, scoped validity, and contestable repair as proposed accountability tests that were not implemented or evaluated.}
  \label{fig:pipeline}
\end{figure*}

Figure~\ref{fig:pipeline} synthesizes the observed chain and the proposed accountability overlay. It separates the contextual public corpus from the Studio pipeline and distinguishes observed transformations from design propositions that remain to be implemented and evaluated.

\subsection{Accountable translation is a normative property of the pipeline}\label{accountable-translation-is-a-normative-property-of-the-pipeline}

``Prompt-to-app'' is an inadequate unit for this case. An explicit domain requirement entered a compiled contract, was assigned to an app family, appeared in a plan and possibly a static artifact, encountered analyzer and package checks, and reached---or did not reach---an organizational registry state. \textbf{Translation} names that observable succession of representations. \textbf{Accountable translation} names the desired property of the succession, not the mere fact that it occurred.

We define a translation as accountable when consequential changes to professional commitments are: \textbf{attributable}, so the stage, template, model, rule, or reviewer responsible is locatable; \textbf{inspectable}, so the domain author can compare representations; \textbf{validated within an explicit scope}, so a local check cannot silently stand for end-to-end quality; and \textbf{contestable}, so an author can accept, edit, or reject an interpretation through domain-legible repair. The accountable objects in this study are purpose and interaction, decision rights, evidence and data boundaries, validity scope, and repair rights.

In education, preserving a topic while changing who acts, who judges, or what evidence returns is a redistribution of professional authority, not a neutral paraphrase.

\subsection{Agentic authoring requires scoped status claims}\label{agentic-authoring-requires-scoped-status-claims}

The four observed layers are not interchangeable forms of one validity construct. Pipeline-step completion is bookkeeping about termination and persistence. A package/security check evaluates specified technical contracts. Analyst-assessed alignment is an interpretive comparison of explicit brief requirements with inspected static material. Registry presence is an organizational snapshot. A numeric score from one layer cannot truthfully certify the others.

T1 and T2 stored regex-enhanced package/security scores of 95 against minimum threshold 60 without establishing score calibration, runtime behavior, author-endorsed alignment, human review, or registry availability. T3 reached completed ledger states without a usable payload. Valid JSON analyzer records contained QA notes requesting manual review while later package checks stored 95. These observations are not inherently contradictory if every subsystem reports a narrow scope; the problem arises when provenance or interface language leaves that scope unclear.

The corresponding design proposition is to label every status with its evidence and residual uncertainty. For these records, a package result could state: ``The regex-enhanced package/security check stored score 95 against a minimum threshold of 60; the score's calibration, runtime behavior, brief alignment, human review, and registry publication remain unverified.'' An alignment view could show which explicit requirements were found, transformed, not found, or unresolved, without presenting analyst judgment as pedagogical effectiveness. Availability should state the snapshot, pending actor, and next authorized action. Such separation supports appropriate reliance without treating a locally correct check as globally sufficient.

Check timing is itself an interaction problem: recent work on multi-step agentic tasks frames confirmation frequency as a trade-off between interruption burden and downstream recovery cost \citep{zhou2026confirmation}. Pedagogical correspondence is partly interpretive, however, so when and what educators should confirm remains an empirical question rather than a fixed checkpoint schedule.

\subsection{Proposition: semantic diffs can make handoffs inspectable}\label{proposition-semantic-diffs-can-make-handoffs-inspectable}

Agentic pipelines can produce extensive internal traces, but the present cases do not establish what participants could see or understand. We therefore treat \textbf{handoff observability} as a design proposition: reorganize trace information around questions a domain expert can ask, then evaluate the proposal with users.

\begin{itemize}
\tightlist
\item
  What did the compiler add, infer, or replace in my brief?
\item
  Which app-family template shaped the plan, and why was it selected?
\item
  Which requested actions and controls appear in the generated interaction?
\item
  What did the analyst reject, and what did the package checker actually report?
\item
  What changed after my revision or exact retry?
\item
  What remains between a validated draft and an available app?
\end{itemize}

A \textbf{semantic diff} would compare domain-relevant representations, not only text or code. Before generation, an interface could present an educator-editable contract containing learning purpose, learner action, AI action, prohibited action, professional control, evidence returned, data boundary, and success condition. After planning and generation, it could show each item as preserved, transformed, not found, unresolved, or introduced, with links to supporting plan and screen evidence. Applied retrospectively to T2, such a view would have contrasted originality dialogue with the inspected recall workflow even though the title remained aligned. This is an analytic illustration, not evidence that the interface would improve authoring.

Pail and Jelly demonstrate editable intermediate representations, SimStep exposes task-level abstractions in educational authoring, and semantic guidance bridges intent and generated UI \citep{zamfirescu2025pail,cao2025jelly,kaputa2026simstep,park2026semantic}. Multi-agent debugging further shows that visibility becomes useful when coupled with localization and reversible intervention \citep{epperson2025debugging}. Our cases extend the target from local design decisions or one generated interface to professional commitments, status evidence, authority, and availability across heterogeneous technical and organizational gates.

Cross-agent provenance could connect a compiler decision to the plan it shaped, the component that implemented it, the check that inspected it, and the person or policy authorized to approve it. This would make responsibility locatable and allow a changed requirement to invalidate only affected downstream decisions. Whether such provenance is comprehensible, useful, or burdensome remains an empirical question.

\subsection{Proposition: repair should be domain-legible and contestable}\label{proposition-repair-should-be-domain-legible-and-contestable}

T3 shows that withholding an unverified payload can coexist with stored guidance that does not specify a domain-level change. The coder refused malformed or unsafe dynamic payloads and refunded credits. The accompanying fidelity terms belonged to a platform-specific template grammar. ``Reflection journal'' and ``artifact review'' did not state how the labels related to adaptive revision, misconception feedback, or a learner dashboard. The trace does not show whether the account holder read or understood these messages, so the claim is about the messages' content, not user experience.

A repair interface could translate in both directions. For every failed contract term, it could show: (1) the relevant original statement; (2) the missing or conflicting visible interaction; (3) a proposed change in domain language; and (4) consequences for data, control, and scope. A hypothetical message might say: ``Your brief asks learners to revise misconceptions over time. The current plan has no screen where they record a revised explanation. Add a revision step visible to learner and educator, or remove longitudinal revision from the success criteria.'' The author could accept, edit, or reject the interpretation. This proposal remains unimplemented and unevaluated.

This two-way translation treats repair as procedural contestability: the author can understand the interpretation, locate responsibility, and act at the relevant point in the pipeline rather than appeal only after a final outcome \citep{yurrita2025contestability}.

A shared vocabulary could also expose variation before another build. An exact-retry control could state whether the system will reuse the compiled contract, refresh model output, change routing, or apply new defaults. Repairability would then concern control over which representation is being repaired, not only clearer error wording. The T3 sequence motivates this proposition but cannot establish its effectiveness.

\subsection{A handoff accountability contract}\label{a-handoff-accountability-contract}

The findings can be operationalized as a contract at each transition. A trace would record: (1) the source professional commitment; (2) the transformation made; (3) the responsible model, template, rule, or reviewer; (4) the evidence and scope supporting the resulting status; (5) the available repair or contestation path; and (6) the consequence for downstream availability. This contract is a design proposition, not an implemented or evaluated feature.

The framework therefore has one hierarchy rather than four competing lists. A professional commitment is examined across five accountability objects: purpose and interaction, decision rights, evidence and data boundaries, validity scope, and repair rights. Each consequential handoff is then tested for attributable change, inspectability, explicitly scoped validity, and contestable repair. The six contract fields operationalize those four tests as trace records. In this setting, the chain is accountable to the domain author and authorized reviewer, for transformations that alter any of the five objects, through a remedy that permits an interpretation to be inspected, accepted, edited, rejected, or retried before downstream approval or availability.

An interface could apply it by eliciting users, domain problem, AI and professional actions, evidence, data boundary, safeguards, and success; showing a semantic diff across brief, compiled contract, plan, and artifact; reporting pipeline completion, package integrity, professional review, and availability separately; translating validator categories into editable domain-level repairs; and stating who can move an artifact through review to registry. Oversight would then become an interaction architecture with legible responsibility and meaningful contribution rather than a final checkbox \citep{faas2026oversight}. Transfer beyond education remains a hypothesis requiring study under each domain's commitments and approval regime.

\subsection{Alternative explanations delimit the interpretation}\label{alternative-explanations-delimit-the-mechanism}

Accountable translation does not imply that every transformation is harmful or that verbatim obedience is desirable. Accessibility, storage, review, and error-state requirements filled genuine specification gaps. The coder withheld T3's unverified payloads from the subsequent artifact path. T1 partially preserved its non-answering boundary, T2 preserved a relevant identity, and T3's structured rewrite changed capability extraction. The normative question is whether consequential substitutions and constraints are visible and contestable, not whether the first prompt is always correct.

Several software explanations remain unresolved. Generic workflows could arise from templates, planning, generation, fallbacks, or ordinary defects; repeated-output variation could reflect model or system state. The score-95 records identify a regex-enhanced package/security checker and minimum threshold 60, but the score's calibration and relation to runtime or semantic quality remain unverified. Instrumentation failures may be separate from the build path, and static absence of an API reference does not rule out runtime injection. T2's requested AI-contribution judgment may also have been infeasible or intentionally constrained.

Context also matters. Access disruption and limited time may have selected the accounts reaching Studio and encouraged short reformulations or repeats. Public chat favors concise writers and reporters; unposted deliberation is absent. Registry absence may mean publication was never attempted or review remained pending. Insider involvement may have shaped our correspondence judgments. These alternatives prevent causal and prevalence claims while delimiting the representation and status differences that future interfaces should test.

\section{Limitations}\label{limitations}

This is a bounded critical trace study, not a prevalence, usability, or effectiveness evaluation. The public corpus contains 37 units from 23 display names, but display names are not verified people and contributors are not the workshop attendance denominator. Group framings represent collective work reported by one account and cannot be interpreted as that reporter's individual attitude. The 18 later ideas show what was posted after a scaffold, not individual improvement. Access disruption and public-chat compression shaped what was preserved.

The Studio analysis contains six attempts from three accounts, four from T3. This supports process tracing and negative-case analysis, not prevalence estimates. Stored flags determined eligibility, but their assignment provenance was incomplete. We did not link Studio cases to public ideas, execute generated apps, interview account holders, observe noticed interface states, or evaluate classroom use. Static inspection may miss runtime behavior; platform scores are implementation-specific, and their calibration against runtime or semantic quality was not established.

The insider roles described above limit interpretive independence. First-pass public unit assignments were not preserved; the Codex-assisted, aggregate-aware second pass was non-independent and offers sensitivity analysis rather than corroboration. The team also assessed brief--artifact correspondence. Future work should preserve unit matrices, use independent domain reviewers, capture responses to semantic diffs and repair, compare compiler representations, and follow artifacts through review and permitted use.

Provider selection, inference settings, complete software-version provenance, and the exact timing or intent of registry submission were unavailable. Consequently, we cannot attribute observed variation to a model, compiler, validator, or author action, nor treat registry absence as a failed publication attempt. The empirical figures summarize bounded records and analyst decisions; they do not show runtime behavior, participant interpretation, or an evaluated interface.

Finally, the study concerns educational authoring in one professional-learning context. Its proposed transferable contribution is an analytic lens and vocabulary for studying accountability across semantic and organizational handoffs---not a universal account of educators, educational systems, or agentic authoring.

\section{Conclusion}\label{conclusion}

Natural-language authoring did not collapse professional intent directly into an app. In six eligible Studio attempts, briefs passed through compiled platform contracts, application-family assignments, plans, generated material, scoped checks, repair messages, and an organizational registry. The traces showed both governance accretion and interaction reshaping. Two payloads stored regex-enhanced package/security scores of 95 against minimum threshold 60 while analyzer reservations and unresolved brief correspondence remained; four attempts in one account produced no usable payload. These records do not establish score calibration, runtime behavior, account-holder experience, or classroom effectiveness. They show that local technical status, semantic correspondence, repair, and availability must be treated as distinct claims.

For HCI, this shifts evaluation from prompt quality to the accountability of a delegated transformation chain. Accountable translation asks whether consequential changes to purpose, decision rights, evidence boundaries, validation scope, and repair rights remain attributable, inspectable, explicitly scoped in validation, and contestable. The framework yields testable design propositions: semantic diffs that expose change, status labels tied to their evidence, provenance that locates responsibility, and repair controls written in domain language. These propositions remain to be built and evaluated. Without them, widening access to software creation may also widen the distance between the professional held responsible for an application and the decisions embedded in it.

\begin{acks}
This work was supported by the AIEDx grant from the Institute of AI in Education (IAIED), Singapore (UEN 53487769M).
This work acknowledges funding support from the Design-centric Interdisciplinary Creative Problem Solving (ICPS-D): Developing Critical and Creative Skills in Teachers project.
\end{acks}

\section{Generative AI Assistance Disclosure}

The authors used OpenAI Codex to assist with coding de-identified analytic units, aggregate consistency checking, manuscript restructuring, and prose drafting. Codex was not treated as an independent coder: aggregate first-pass counts were visible during the second pass, unit-level first-pass assignments were unavailable, and no reliability statistic is claimed. Human authors defined the research questions and analytic boundaries, selected the argument, and remain accountable for the work's accuracy, originality, and integrity. The reported descriptive counts and interpretations should be read in light of this model-assisted analytic procedure.

\bibliographystyle{ACM-Reference-Format}
\bibliography{refs}

\end{document}